\documentclass{raa}

\usepackage{graphicx,times}             %for PS/EPS graphics inclusion, new
\usepackage{amssymb}

\begin{document}

   \title{Grids of rotating stellar models with masses between 1.0 and 3.0 $\mathrm{M}_{\odot}$
%\footnotetext{$*$ Supported by the National Natural Science Foundation of China.}
}
%   \subtitle{I. Place Your Subtitle Here}

   \volnopage{Vol.0 (200x) No.0, 000--000}      %%preserved for Editor. DOn't remove!
   \setcounter{page}{1}          %%starting page, preserved for Editor. DOn't remove!

   \author{Wu-Ming Yang
      \inst{1,2}
   \and Shao-Lan Bi
      \inst{1}
   \and  Xiang-Cun Meng
      \inst{2}
   }

   \institute{Department of Astronomy, Beijing Normal University, Beijing 100875,
China; {\it yangwuming@ynao.ac.cn; yangwuming@bnu.edu.cn}\\
  \and
    School of Physics and Chemistry, Henan Polytechnic University,
Jiaozuo 454000, Henan, China.\\
   }

   \date{ }

\def\dsm{$\mathrm{M}_{\odot}$}

\abstract{We calculated a grid of evolutionary tracks of
rotating models with masses between 1.0 and 3.0 \dsm{} and a resolution $\delta M \leq 0.02$ \dsm{},
which can be used to study the effects of rotation on stellar evolutions and on the characteristics
of star clusters. The value of $\sim$2.05 \dsm{} is a critical mass for the effects of rotation
on stellar structure and evolution. For stars with $M >$ 2.05 \dsm{}, rotation leads
to an increase in the convective core and prolongs the lifetime of main sequence (MS);
rotating models evolve slower than non-rotating ones; the effects of rotation on the evolution
of these stars are similar to those of convective core overshooting. However for stars with 1.1
$<$ M/\dsm{} $<$ 2.05, rotation results in a decrease in the convective core and
shortens the lifetime of MS; rotating models evolve faster than non-rotating ones.
When the mass is located in the range of $\sim$1.7 - 2.0 \dsm{}, the mixing caused by rotationally
induced instabilities is not efficient; the hydrostatic effects dominate the effect on the evolution of
these stars. For the models with masses between about 1.6 and 2.0 \dsm{}, rotating models always
exhibit lower effective temperatures than non-rotating ones at the same age during the MS stage.
For a given age, the lower the mass, the smaller the change in the effective temperature.
Thus rotations could lead to a color spread near the MS turnoff in the color-magnitude
diagram for the intermediate-age star clusters.
}

   \authorrunning{Yang et al.}            %author_head in even pages
   \titlerunning{Grids of rotating stellar models}  % title_head in odd pages

  \maketitle

\section{Introduction}           %% first-level sections will be auto-capitalized
%\label{sect:intro}

Recently, it has been discovered that some intermediate-age star clusters have a double
or extended main-sequence turnoffs (MSTOs) in their color-magnitude diagrams (CMDs)
(Mackey \& Broby Nielsen \cite{mack07}, Mackey et al. \cite{mack08}, Milone \cite{milo09},
Goudfrooij et al. \cite{goud09, goud11}). Platais et al. (\cite{plat12})
also found that the upper main sequence of the open cluster Trumpler 20 with an age of
about 1.3 Gyr appears to show an enlarged color spread which is not normally seen in the CMD
for open clusters in this age group. The double or extended MSTOs are
interpreted as that the star clusters have two or multiple stellar populations with similar
metal abundance but with differences in age of $\sim$200 - 300 Myr (Mackey et al. \cite{mack08}),
which is contrary to the traditional knowledge that a star cluster is comprised of stars
belonging to a single, simple stellar population with a uniform age and chemical composition.
However, Platais et al. (\cite{plat12}) also pointed out that the enlarged color spread of Trumpler 20 may
be due to differential reddening. In order to understand the extended MSTO, many scenarios were
proposed by many investigators (Mackey \& Broby Nielsen \cite{mack07}, Bekki \& Mackey \cite{bekk09},
Goudfrooij et al. \cite{goud09}, Bastian \& de Mink \cite{bast09}, Rubele et al. \cite{rube10, rube11},
Yang et al. \cite{yang11b}, Girardi et al. \cite{gira11}). In these scenarios, Yang et al. (\cite{yang11b})
found that the interactions in binary systems can reproduce an extended MSTO of star clusters. However,
the fraction of the interactive binary systems is too low to completely explain the observed features
in star clusters. Another interpretation proposed by Rubele et al. (\cite{rube10, rube11}) and
Girardi et al. (\cite{gira11}) is continuous star formation, lasting $\sim$300 Myr or longer.
However, Platais et al. (\cite{plat12}) noted that `this long period of star formation seems to
be odds with the fact that none of the younger clusters are known to have such a trait', and that some
star clusters with extended MSTO might not experience a self-enrichment. Yang et al. (\cite{yang11b})
also noted that the lasting time for explaining the double MSTO and dual red clump of a star cluster
is not consistent.

Rotation is a property that virtually all stars possess. Observations show that main-sequence (MS) stars with
1.3 $<$ M/\dsm{} $<$ 3.0 have a typical value of 160 km $\mathrm{s}^{-1}$ of $v\sin i$ (Royer et al. \cite{roye07})
which corresponds to a period of $\sim$0.5-0.8 day. In the classical theory of stellar evolution, the effects of
rotation on stellar evolutions are always neglected. However, in fact, rotation is one of the key factors
that can change the evolution and all outputs of stellar models (Maeder \& Meynet \cite{maed00}).
Bastian \& de Mink (\cite{bast09}) calculated the evolutions of rotating models with an initial mass
of 1.5 \dsm{} but with different rotation rates using the stellar evolution code described by
Yoon et al. (\cite{yoon06}) in which the effects of rotation on the structure and mixing induced
by rotation are taken into account. They found that rotating models are cooler and fainter than non-rotating
ones for not extreme rotation rates. They incorporated these effects of rotation on stellar evolutions
into the sythesis of CMD of star clusters. They found that stellar rotations can mimic the effect of
multiple populations in star clusters, whereas in actuality only a single population exists.
However, using the Geneva stellar evolution code, Eggenberger et al. (\cite{egge10}) and Girardi et al. (\cite{gira11})
calculated the evolutions of rotating models with an initial velocity on the zero-age
main sequence (ZAMS) of 150 km $\mathrm{s}^{-1}$, they found that rotating models can be slightly hotter
and brighter than non-rotating ones. The isochrone of rotating stars has a slightly hotter and brighter
turnoff with respect to that of non-rotating stars (Girardi et al. \cite{gira11}), which is contrary to
the calculation result of Bastian \& de Mink (\cite{bast09}). The origin of the extended MSTO of intermediate-age star
clusters is still an open question. In order to understand the effects of stellar rotation on the evolutions
of stars and on the CMD of star clusters, the evolutions of rotating stars need to be studied in more detailed.

In this paper, we mainly focus on the effects of rotation on the evolutions of stars,
especially on the evolutions of the intermediate-mass MS stars which corresponds to the MSTO
stars of intermediate-age star clusters. The paper is organized as follows:
we give a physical description of the effects of rotation in Section 2,
present the calculation results in Section 3 and discuss and summarize them in Section 4.

\section{Physical description of rotation}
\subsection{Effects of rotation on stellar structure and evolution}
The equations of stellar structure of a rotating star were built up by Kippenhahn \& Thomas (\cite{kipp70})
and modified by Endal \& Sofia (\cite{enda76}) and Meynet \& Maeder (\cite{meyn97}) to calculate shellular rotation.
Rotation affects stellar structure and evolution mainly in four ways:

1. The first one is the effect of centrifugal force on the effective gravity. This effect can be
considered directly in the equation of hydrostatic equilibrium.

2. The second is that the equipotential or isobar surfaces are no longer spheres for
a rotating star because the centrifugal force is always perpendicular to the axis of rotation
and is not, in general, parallel to the force of gravity. Thus the assumption of spherical symmetry
for non-rotating stars is no longer fit for rotating ones. A rotating star may be a spheroid.
This leads to the fact that almost all of structure equations are affected and the independent
variable of equations of stellar structure need to be redefined.

3. The third is that the radiative flux is not constant on an isobar surface
because the radiative flux varies with the local effective gravity, i.e. the Von Zeipel effect
(Von Zeipel \cite{zeip24}), which affects the radiative equilibrium equation and the instability
of convection by changing the radiative temperature gradient.

4. The final results from the transport of angular momentum and the mixing of elements
caused by rotationally induced instabilities. The mixing happening in the stellar interiors can affect
the density distribution by changing the mean molecular weight. Thus it can affect the radius of stars,
the instability of convection, and so on.

The first three effects are directly incorporated into the equations of stellar structure (Endal \& Sofia
\cite{enda76}). In a nonrotating model the independent variable, $M_{r}$, is the mass contained within a spherical surface.
However, in the rotating one the independent variable, $M_{p}$, is the mass contained within an isobar surface.
The algorithmic methods of the isobar surface and physical quantities on the isobar surface can be found
in Endal \& Sofia (\cite{enda76}), Meynet \& Maeder (\cite{meyn97}), and Yang \& Bi (\cite{yang06}) in detail.

\subsection{Angular momentum loss, angular momentum transport and the mixing of elements}

In this context, we assumed that angular momentum loss is caused by magnetic stellar winds
and the loss happens only when the stellar envelope is convective. The parameterized Kawaler's relations
(Kawaler \cite{kawa88}, Chaboyer et al. \cite{chab95})
\begin{equation}
 \frac{dJ}{dt}= f_{k}K_{w}\frac{R}{R_{\odot}}^{2-\beta}\left( \frac{M}{M_{\odot}} \right)^{-\beta/3}
\left( \frac{\dot{M}}{10^{-14}} \right)^{1-2\beta/3}\Omega^{1+4\beta/3}
~~~~~~(\Omega < \Omega_{crit}),
\label{eqw1}
\end{equation}
\begin{equation}
 \frac{dJ}{dt}= f_{k}K_{w}\left( \frac{R}{R_{\odot}} \right)^{2-\beta}\left( \frac{M}{M_{\odot}} \right)^{-\beta/3}
\left( \frac{\dot{M}}{10^{-14}} \right)^{1-2\beta/3}\Omega\Omega_{crit}^{4\beta/3}
 ~~~~~~(\Omega \geq \Omega_{crit})
\label{eqw2}
\end{equation}
were used to calculate the rate of angular momentum loss, where $f_{k}$, $\beta$, and $\Omega_{crit}$ are adjustable
parameters, $K_{w} = 2.035\times10^{33}\times(1.442\times10^{9})^{\beta}$ in cgs units, the mass-loss rate $\dot{M}$
is set to $2.0\times10^{-14}$ \dsm{} $\mathrm{yr^{-1}}$. The value of $f_{k}$ calibrated to the Sun is 2.5,
however, which may overestimate the rate of angular momentum loss for stars with mass larger than the Sun
because these stars have a shallower convective envelope than the Sun.
Thus we adopted a value of 1.0 for the stars with mass larger than the Sun.

\begin{table}
\begin{center}
\caption[]{The values of the parameters for angular momentum loss,
the transport of angular momentum, and element mixing.}
\label{tabpara}
 \begin{tabular}{cccccc}
  \hline\noalign{\smallskip}
Parameter &   $f_{k}$ & $\beta$ & $\Omega_{crit}$  &  $f_{\Omega}$ & $f_{c}$   \\
  \hline\noalign{\smallskip}
Value  & $1.0^{a}(2.5^{b})$  &  1.5   & 5$\Omega_{\odot}$  &  1.0 & 0.2 \\
  \noalign{\smallskip}\hline
\end{tabular}
   \begin{list}{}{}
     \item Note.-- $a$ for models with $M >$ 1.0 \dsm{}; $b$ for 1.0 \dsm{} model.
   \end{list}
\end{center}
\end{table}

In addition, we assumed that the rotation is uniform on the ZAMS and in all convective regions of a star.
The angular momentum loss, expansion and/or contraction of stars lead to the occurrence of
differential rotation in radiative regions. The rotational instabilities induced by rotation
not only can transport angular momentum but also can mix chemical elements.
Because the horizontal turbulent is stronger than the vertical turbulent, the angular velocity
and chemical compositions are constant on an isobar (Zahn \cite{zahn92}). Thus we only consider
the vertical transports of angular momentum and chemical elements which obey two coupled nonlinear
diffusion equations:
\begin{equation}
 \rho r^{2}\frac{I}{M}\frac{\partial\Omega}{\partial t}=
 f_{\Omega}\frac{\partial}{\partial r}(\rho r^{2}\frac{I}{M}D\frac{\partial\Omega}{\partial r})
 -\frac{\partial}{\partial r}(\rho r^{2}\frac{I}{M}\Omega \dot{r})
\label{eqt1}
\end{equation}
for the transport of angular momentum and
\begin{equation}
 \frac{\partial X_{i}}{\partial t}=(\frac{\partial X_{i}}{\partial t})_{nuc}
 -\frac{1}{\rho r^{2}}\frac{\partial}{\partial r}(\rho r^{2}X_{i}V_{i})
 +f_{\Omega}\frac{1}{\rho r^{2}}\frac{\partial}{\partial r}(\rho r^{2}f_{c}D\frac{\partial X_{i}}{\partial r})
\label{eqt2}
\end{equation}
for the change in the the mass fraction $X_{i}$ of species $i$, where the fraction $I/M$ is the momentum of
inertia per unit mass, and $D$ is the diffusion coefficient due to the rotational instabilities 
including meridional circulation, the Goldreich-Schubert-Fricke instability 
(Goldreich \& Schubert \cite{gold67}; Fricke \cite{fric68}), and the secular shear instability.
The criterion and the estimate of the coefficient $D$ of these instabilities were summarized 
by Endal \& Sofia (\cite{enda78}) and Pinsonneault et al. (\cite{pins89}) in detail. The second term
on the right-side of Eq. (\ref{eqt1}) describes the change of angular velocity induced by contraction and/or
expansion of stars. The first and second term on the right-side of Eq. (\ref{eqt2}) is due to the nuclear reaction
and gravity settling diffusion, respectively. The velocity, $V_{i}$, of the gravity settling diffusion is given by
Thoul et al. (\cite{thou94}). The adjustable parameter $f_{\Omega}$ is introduced to represent some inherent
uncertainties in the diffusion equation, and the $f_{c}$ is used to account for the fact that the transports
of angular momentum and a chemical species occur on different timescales and is less than one.
The parameters $f_{\Omega}$ and $f_{c}$ may depend on stellar mass. However, for simplicity, they are 
assumed to be a constant.
The values of the adjustable parameters in Eqs. (\ref{eqw1})-(\ref{eqt2}) are summarized in Table \ref{tabpara}.
The effects of magnetic fields in the stellar interiors are not considered in our models. The magnetic fields
are introduced to explain the flat rotational profile in the Sun (Brown et al. \cite{brow89}, Kosovichev et al.
\cite{koso97}, Eggenberger et al. \cite{egge05}, Yang \& Bi \cite{yang06, yang08}). However, the gravity waves also can
be competent to explain the solar rotation (Zahn et al. \cite{zahn97}, Talon \& Charbonnel \cite{talo05}).
Especially for the intermediate-mass stars which do not experience magnetic braking, our calculations show that 
the rotational instabilities are efficent enough to transport angular momentum in these stars.

\section{Results}
We used the Yale Rotation Evolution Code (YREC7) to compute the evolutions of rotating and non-rotating models
with masses between 1.0 and 3.0 \dsm{}. The code has been updated with recent input physics over the last three
decades (Endal \& Sofia \cite{enda76, enda78}, Pinsonneault et al. \cite{pins89}, Chaboyer et al. \cite{chab95},
Yang \& Bi \cite{yang07}). The new OPAL EOS tables (Rogers \& Nayfonov \cite{roge02}),
OPAL opacity tables (Iglesias \& Rogers \cite{igle96}), and the opacity tables for low temperature provided
by Alexander \& Ferguson (\cite{alex94}) were used. Energy transfer by convection is treated according to
the standard mixing length theory. The value of 1.72 for the mixing-length parameter ($\alpha$) was calibrated
against the Sun. The initial chemical compositions of our models are fixed at $Z_{0}$ = 0.02 and $X_{0}$ = 0.707.
The initial rotation period $P_{0}$ (or initial rotation velocity $V_{0}$) is a free parameter. We mainly computed
the evolutions of the models with $P_{0} =$ 0.74 and 0.50 day, respectively. All models share the same initial
parameters except for mass and rotation period.

\subsection{For the stars with mass less than 1.5 \dsm{}}
%-------------------------------------------------------------
   \begin{figure}
   \centering
   \includegraphics[width=4cm, angle=-90]{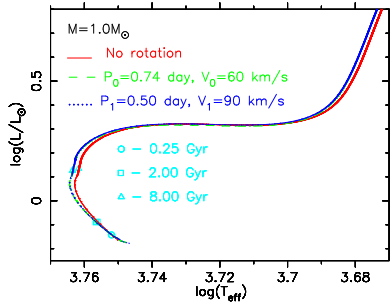}
   \caption{Evolutionary tracks in the Hertzsprung-Russell (HR) diagram for 1\dsm{} models.
   The solid (red) line corresponds to the track of the non-rotating model, the dashed (green)
   and dotted (blue) lines show that of the rotating models with an initial period of 0.74 and 0.50 day,
   respectively. The positions of models at the ages of 0.25, 2.0, and 8.0 Gyr are marked on the tracks.}
   \label{fm1p0}
   \end{figure}

\begin{figure}
   \centering
   \includegraphics[width=4cm, angle=-90]{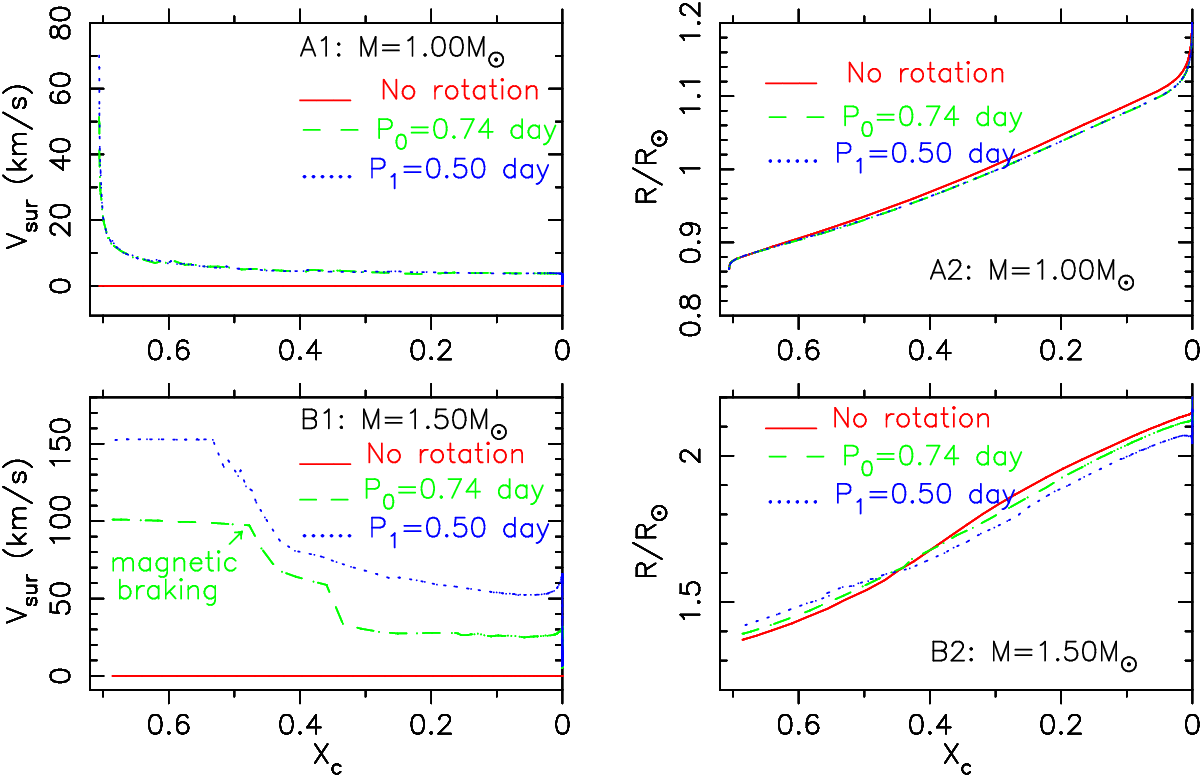}
   \includegraphics[width=4cm, angle=-90]{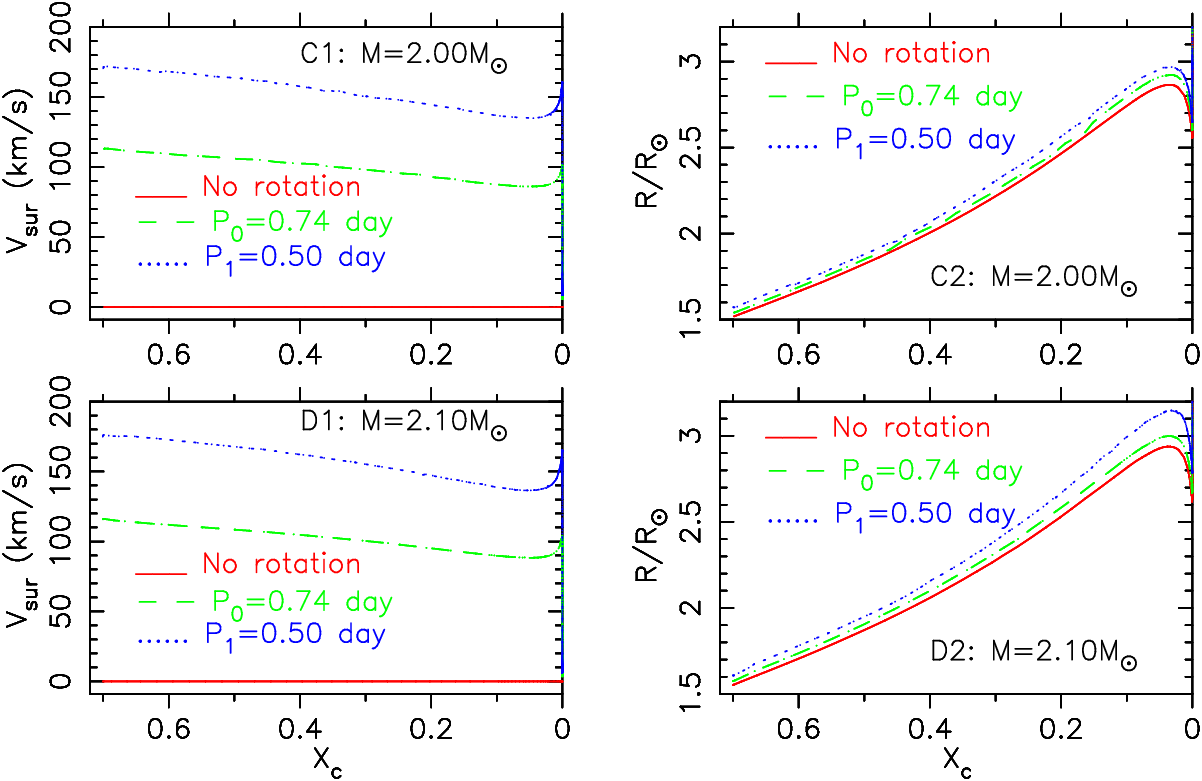}
   \caption{The surface (equatorial) velocity and radius as a function of the mass fraction of central
   hydrogen.The solid (red) line corresponds to the non-rotating model, the dashed (green) and dotted (blue)
   lines indicate the rotating models with an initial period of 0.74 and 0.50 day, respectively.}
   \label{fvsur}
   \end{figure}

Figure \ref{fm1p0} shows the evolutionary tracks of the 1.0 \dsm{} models with and without rotation.
No matter what the initial rotation period, the efficient magnetic braking almost results in the same
rotational velocity (see the panel A1 in Fig. \ref{fvsur}) within a few hundred Myr, which leads to
the fact that the evolutionary tracks of the two rotating models are almost overlapped in the
Hertzsprung-Russell (HR) diagram. When the age of stars is younger than about 0.25 Gyr, the effective
temperature and luminosity of the rotating models are slightly lower than that of the non-rotating ones.
However, when age $>$ 0.25 Gyr, the effective temperature of the rotating models is higher
but luminosity is lower than that of the non-rotating one at the same age.

The rotational mixing takes into effect only after composition gradients have been produced.
When age $<$ 0.25 Gyr, the variation of chemical compositions with respect to radius is very small
for a 1.0 \dsm{} model, thus the effects of rotational mixing can be neglected and the hydrostatic effects
of rotation play a dominant role in this stage. The hydrostatic effects of rotation result in a decrease in
the effective gravity of stars, which can slightly decrease the central temperature and enlarge the radius
of rotating models compared to non-rotating ones. Therefore, the rotating models have a slightly lower luminosity
and effective temperature than non-rotating ones.
%-------------------------------------------------------------
%-------------------------------------------------------------

\begin{figure}
   \centering
   \includegraphics[width=4cm, angle=-90]{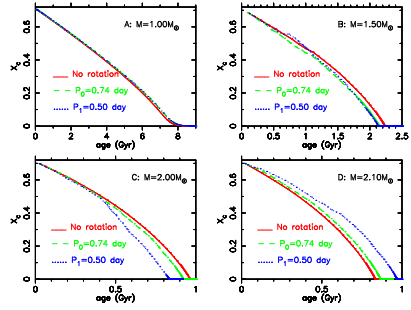}
   \caption{The mass fraction of central hydrogen as a function of age. The solid (red) line corresponds to the non-rotating model,
   the dashed (green) and dotted (blue) lines show the rotating models with an initial period of 0.74 and 0.50 day,
   respectively.}
   \label{fxa}
   \end{figure}

\begin{figure}
   \centering
   \includegraphics[width=4cm, angle=-90]{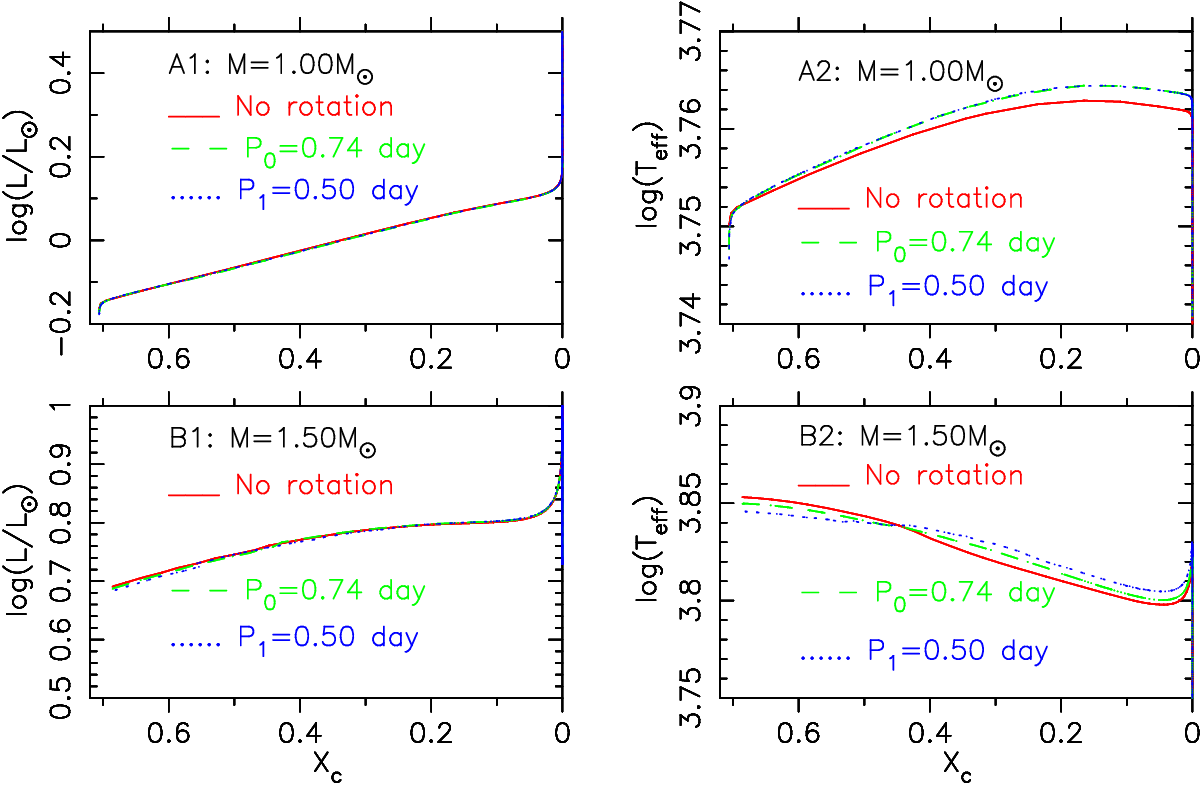}
   \includegraphics[width=4cm, angle=-90]{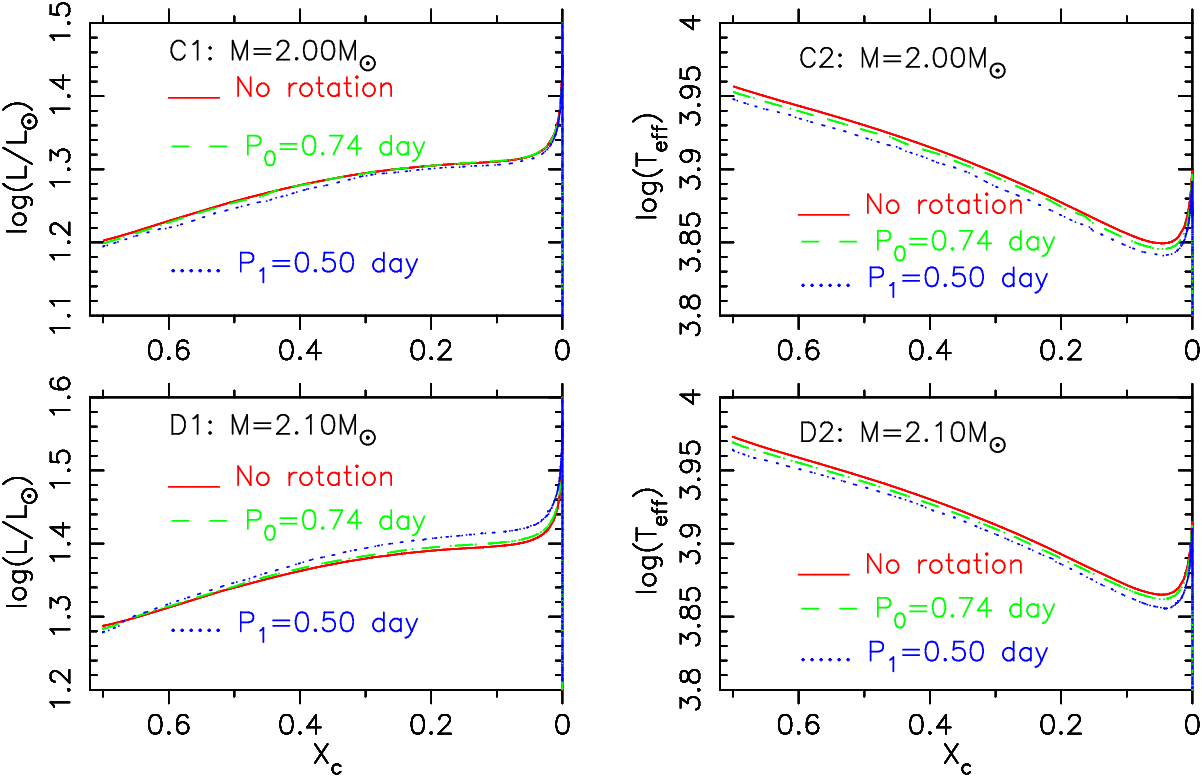}
   \caption{The luminosity and effective temperature as a function of the mass fraction of central
   hydrogen.The solid (red) line corresponds to the non-rotating model, the dashed (green) and dotted (blue)
   lines indicate the rotating models with an initial period of 0.74 and 0.50 day, respectively.}
   \label{fltx}
   \end{figure}

\begin{figure}
   \centering
   \includegraphics[width=4cm, angle=-90]{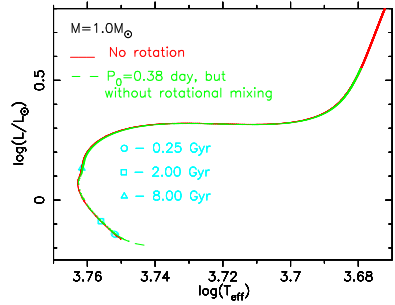}
   \caption{Same as Fig.\ref{fm1p0} but the rotating model without including the effects of rotational mixing.}
   \label{fm1p0b}
   \end{figure}

When age $>$ 0.25 Gyr (the central hydrogen abundance $X_{c} \lesssim$ 0.69), due to the angular momentum loss,
on the one hand, the rotation rate becomes low enough to neglect the influences of the hydrostatic effects
on the radius; on the other hand, the increase in the gradients of angular velocity and chemical compositions
with respect to radius makes that the rotational mixing begins to play a dominant role by feeding more
fresh hydrogen fuel into the hydrogen-burning region and by transporting helium outwards,
which can lead to an increase in the mean density, i.e. decrease in radius, and deceleration of evolution.
Thus the rotating models exhibit a higher central hydrogen abundance (see the panel A in Fig. \ref{fxa})
and effective temperature but slightly lower luminosity than non-rotating ones at the same age.
When the models evolved to the same evolutionary state (the same central hydrogen abundance),
the luminosity of rotating models is only slightly higher than that of non-rotating ones by
consuming more hydrogen, however the radius of rotating models is still smaller than that of
non-rotating ones (see the panel A2 in Fig. \ref{fvsur}) due to the effects of the rotational mixing,
therefore the effective temperature of rotating models is higher than that of non-rotating ones (see
panels A1 and A2 in Fig. \ref{fltx}). If the rotational mixing is not considered, Fig. \ref{fm1p0b} shows
that rotation almost not affects the evolution of a 1.0 \dsm{} model after the age of 0.25 Gyr.

\begin{figure}
   \centering
   \includegraphics[width=4cm, angle=-90]{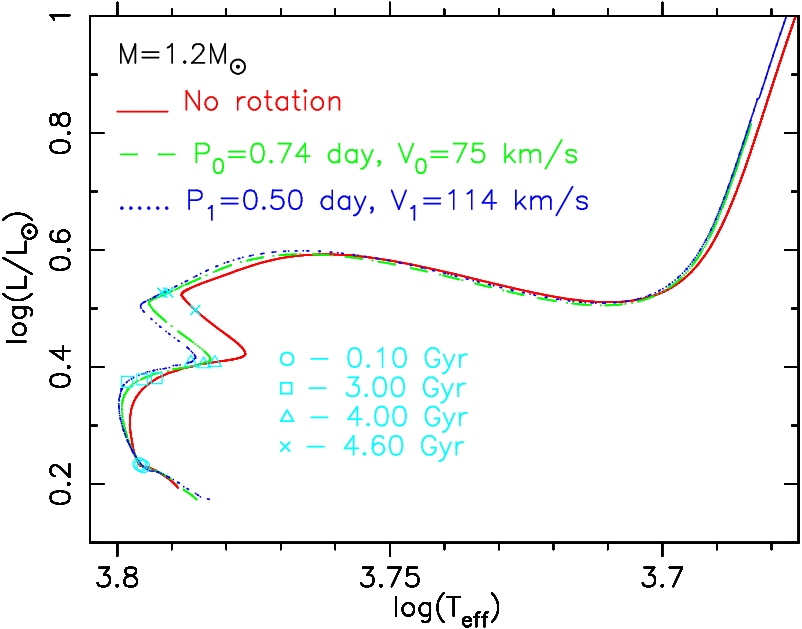}
   \includegraphics[width=4cm, angle=-90]{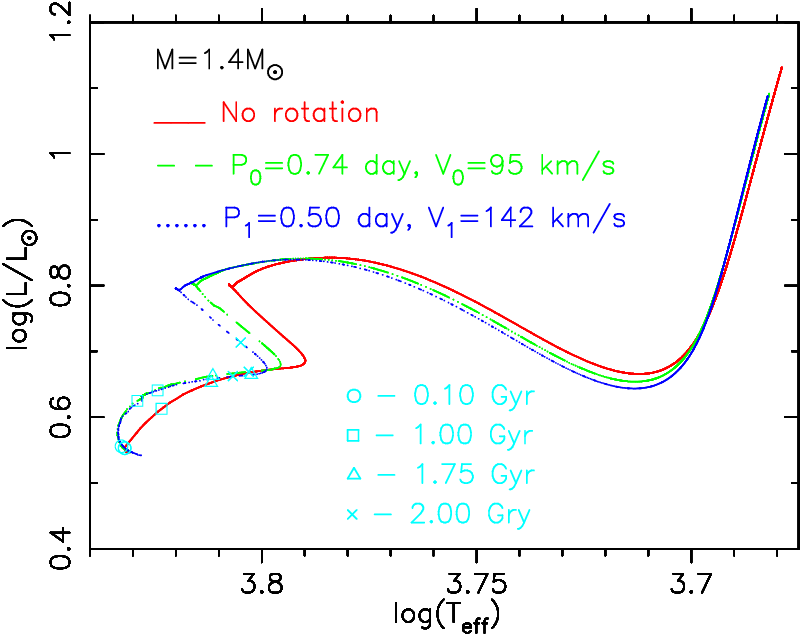}
   \caption{Same as Fig.\ref{fm1p0} but for stars with M = 1.2 and 1.4 \dsm{}, respectively.}
   \label{fig1p2}
   \end{figure}

In Fig. \ref{fig1p2}, we show the evolutionary tracks of rotating and non-rotating models with $M =$
1.2 and 1.4 \dsm{}. The effects of rotation on the evolutions of these stars are similar to those on the
evolution of the 1.0 \dsm{} model. Compared with the tracks of non-rotating models, the tracks of rotating
models move to the left of the HR diagram. For stars with $M \lesssim$ 1.4 \dsm{}, due to the efficient magnetic braking,
the hydrostatic effects of rotation play a dominant role only in about 100 Myr. Then, just as the case
of 1.0 \dsm{} models, the rotational mixing begins to play a dominant role. The rotating models exhibit
a higher effective temperature and an approximately equal luminosity compared to the non-rotating ones when they
evolved into the same evolutionary state [for instance the beginning of the MS hook (Yang et al. \cite{yang11a})
in the HR diagram]. Moreover, for the star with $M=$ 1.4 \dsm{}, Fig. \ref{fig1p2} displays that the rotating models
have higher effective temperatures and luminosities at the age of 1.0 Gyr but have lower effective temperatures
at the age of 1.75 Gyr than non-rotating ones, and that the rotating models evolve faster than the non-rotating ones,
which are different from the evolutions of the models with $M=$ 1.0 \dsm{}.

\begin{figure}
   \centering
   \includegraphics[width=8cm, angle=-90]{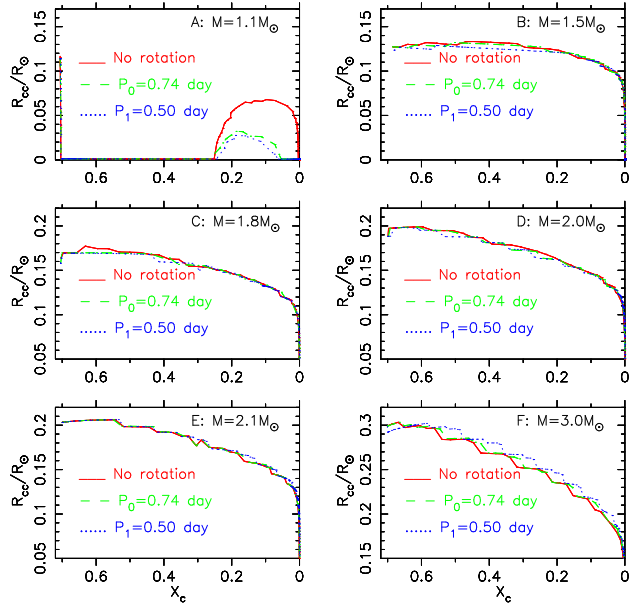}
   \caption{The radius of convective core as a function of the mass fraction of central hydrogen abundance.
   The solid (red) line corresponds to the non-rotating model, the dashed (green) and dotted (blue) lines
   indicate the rotating models with an initial period of 0.74 and 0.50 day, respectively.}
   \label{frc}
   \end{figure}

The radiative temperature gradient can be changed by Von Zeipel (\cite{zeip24}) effect, 
while the adiabatic temperature gradient can be influenced by the mixing of elements caused by 
hydrostatic instabilities. Thus the instability of convection can be affected by the effects of 
rotation. Our calculations show that the effects of rotation can result in a decrease in the radius or mass
of the convective core for stars with 1.1 \dsm{} $\lesssim M <$ 2.05 \dsm{} but an increase for stars with
$M >$ 2.05 \dsm{} (see Fig. \ref{frc}). The decrease in the convective core leads to the fact that
the hydrogen abundance in the convective core of rotating models decreases faster than that of non-rotating ones.
Although the rotational mixing can bring hydrogen fuel into the core from outer layers, which can not compensate
the decrease of the central hydrogen caused by the decrease in the convective core. As a consequence,
the rotating models evolve faster than non-rotating ones during the MS stage for stars
with 1.1 \dsm{} $\lesssim M \lesssim$ 2.0 \dsm{} (see Fig. \ref{fxa}).
Furthermore, Fig. \ref{frc} shows that the faster the rotation, the smaller the convective core,
thus the faster the evolution of the MS stage for stars with 1.1 \dsm{} $\lesssim M \lesssim$ 2.0 \dsm{}.
We labelled this phenomenon as `rotation acceleration'. The rotation acceleration leads to the fact that the rotating
models with $M=$ 1.4 \dsm{} exhibit the higher luminosities at the age of 1.0 Gyr and lower effective temperatures
at the age of 1.75 Gyr than the non-rotating ones (see Fig. \ref{fig1p2}).

%-------------------------------------------------------------
%-------------------------------------------------------------
%-------------------------------------------------------------
%-------------------------------------------------------------

%-------------------------------------------------------------

\subsection{For the stars with masses between 1.5 and 1.7 \dsm{}}

\begin{figure}
   \centering
   \includegraphics[width=4cm, angle=-90]{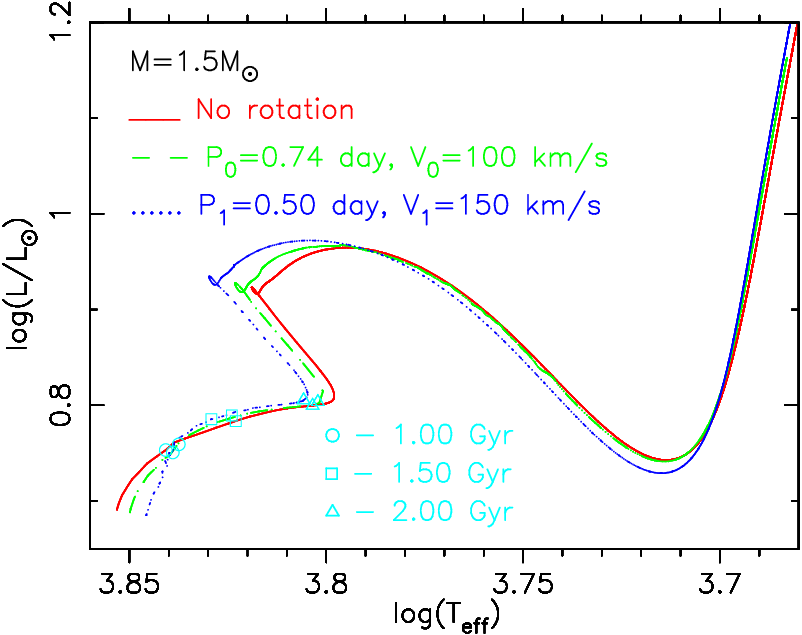}
   \includegraphics[width=4cm, angle=-90]{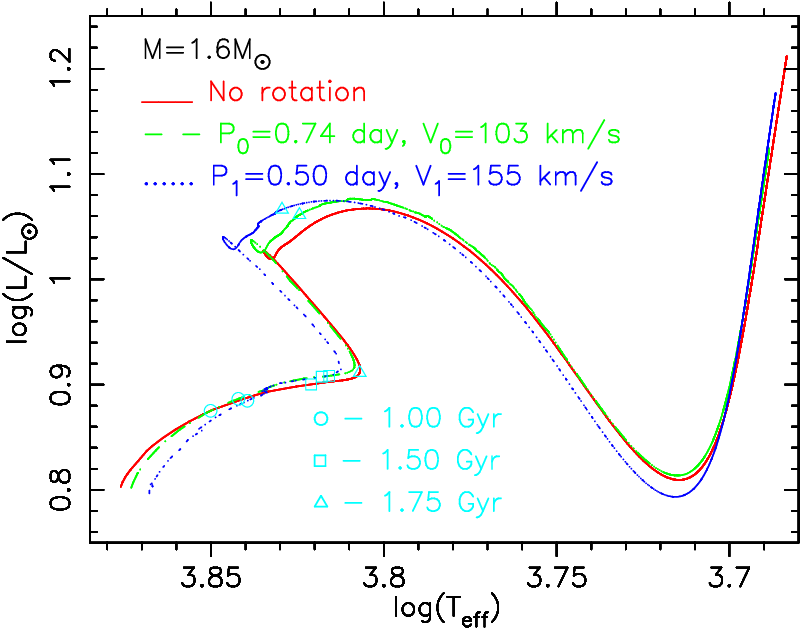}
   \includegraphics[width=4cm, angle=-90]{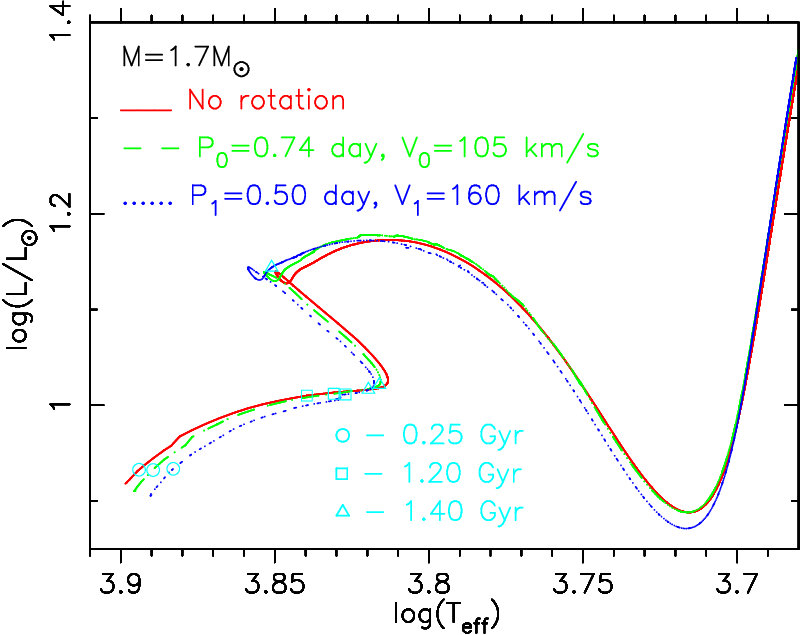}
   \caption{Same as Fig.\ref{fm1p0} but for stars with M = 1.5, 1.6, and 1.7 \dsm{}, respectively.}
   \label{fig1p5}
   \end{figure}

Figure \ref{fig1p5} shows the evolutionary tracks of the rotating and non-rotating models with $M =$ 1.5, 1.6,
and 1.7 \dsm{}. For stars with $M \gtrsim$ 1.5 \dsm{}, their envelopes are radiative at the beginning of MS.
Before the appearance of convective envelope (i.e. before magnetic braking), the stars maintain a fast rotation
(see the panel B1 in Fig. \ref{fvsur}), the hydrostatic effects of rotation dominate the corrections to models
and lead to the rotating models having a larger radius (see the panel B2 in Fig. \ref{fvsur}) and slightly lower
central temperature than the non-rotating ones, thus the rotating models have a slightly lower luminosity and
effective temperature than the non-rotating ones at the same evolutionary stage in the early stage of MS
(see the panels B1 and B2 in Fig. \ref{fltx}). However due to the effects of the rotation acceleration,
as the evolution proceeds, the rotating models can exhibit a slightly higher luminosity than non-rotating one
at the same age.

After the convective envelope has developed, due to the angular momentum loss, on the one hand,
the rotation of the stars begins to become slow, which results in a reduction in the hydrostatic
effects of rotation (for example stellar radius is almost no longer affected by the hydrostatic
effects of rotation); on the other hand, the gradient of angular velocity with respect to radius increases,
which enhances the effects of the rotational mixing. As the evolution proceeds, the rotational mixing
begins to play a dominant role, thus the radius of rotating models changes from larger to smaller than
that of non-rotating ones (see the panel B2 in Fig. \ref{fvsur}). In addition, when the models with
and without rotation evolved to the same evolutionary state of MS stage, their luminosities are
approximately equal, therefore the effective temperatures of rotating models change from lower to higher than
that of non-rotating ones (see the panel B2 in Fig. \ref{fltx}). However, before the MS hook of stars
with $M\gtrsim$ 1.6 \dsm{}, due to the facts that the magnetic braking occurs in the late stage of the
MS and the rotation accelerates the evolutions of rotating models, the effective temperature of rotating models
is always lower than that of non-rotating ones at the same age. For example,
at the age of 1.5 Gyr, the effective temperature of the star with $M=$ 1.6 \dsm{} is about 6620 K for
the non-rotating model but is 6540 K for the rotating model with $T_{0}=$ 0.74 day.
The difference in the effective temperatures is 80 K.

Moreover, Fig. \ref{fig1p5} shows that the luminosities of rotating models are slightly higher than that
of non-rotating one at the end of MS hook. This is because rotation hardly affects the convective core
when $X_{c}<$ 0.1 for these stars and the rotational mixing makes rotating models consuming slightly more
hydrogen than non-rotating ones during the MS hook.

%-------------------------------------------------------------

%-------------------------------------------------------------
%-------------------------------------------------------------
   \begin{figure}
   \centering
   \includegraphics[width=4cm, angle=-90]{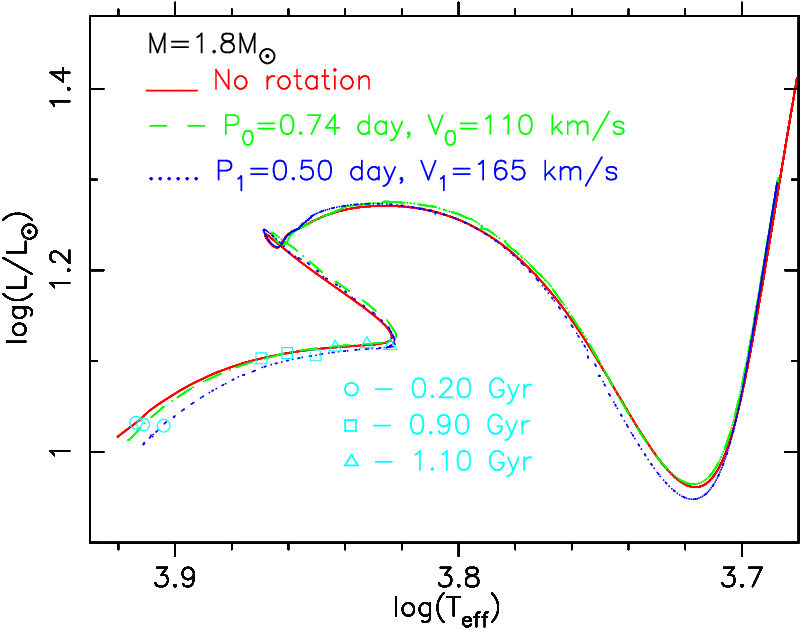}
   \includegraphics[width=4cm, angle=-90]{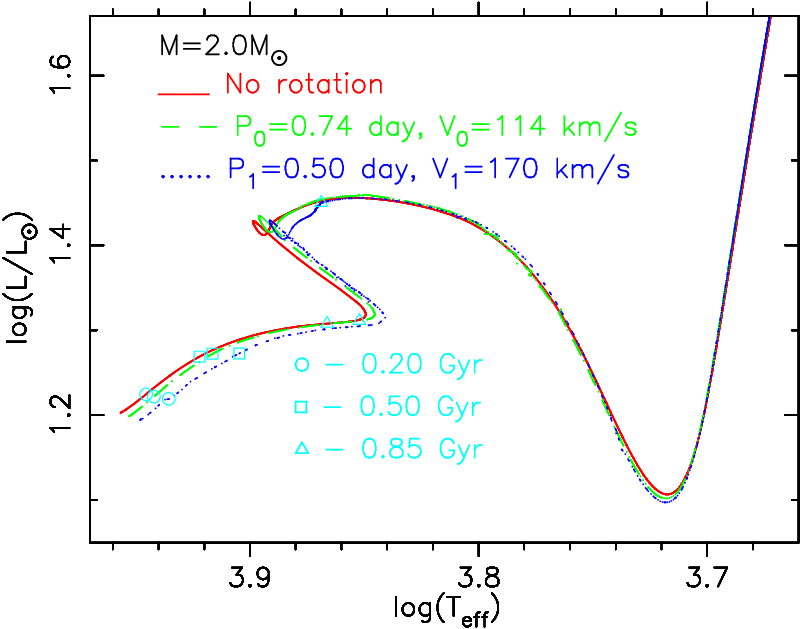}
   \caption{Same as Fig.\ref{fm1p0} but for the models with $M =$ 1.8 and 2.0 \dsm{}, respectively.}
   \label{fig1p8}
   \end{figure}

%-------------------------------------------------------------

\subsection{For the Stars with masses between 1.8 and 2.0 \dsm{}}

The evolutionary tracks of the rotating and non-rotating models with $M=$ 1.8 and 2.0 \dsm{} are shown
in Fig. \ref{fig1p8}. For these stars, the rotating models almost exhibit lower luminosities and effective
temperatures than non-rotating ones when they reached the same evolutionary state in the whole MS stage
(see the panels C1 and C2 in Fig. \ref{fltx}).
Before the MS hook, even at the same age the effective temperatures of the rotating models are also
lower than that of non-rotating ones, but their luminosities are approximately equal.
For example, for the star with $M=$ 1.8 \dsm{}, at the age of 1.1 Gyr the effective temperature and luminosity
is 6970 K and 13.03 $L_{\odot}$ for the non-rotating model but 6670 K and 13.08 $L_{\odot}$ for rotating model
with $T_{0}=$ 0.5 day, respectively. The difference in the effective temperatures is about 300 K.
This is because these stars do not experience magnetic braking during the MS stage.
The fast rotation leads to the fact that the hydrostatic effects of rotation dominate the corrections
to radius. In addition, the effects of the mixing caused by rotationally induced instabilities are partly
counteracted by the effect of the decrease in the convective core in these models.
Thus the rotating models have a larger radius than non-rotating ones at the same evolutionary
state (As a case, see the panel C2 in Fig. \ref{fvsur}). Moreover, the hydrostatic effects of fast rotation also
leads to a decrease in the central temperature. For example, when the star with $M=$ 2.0 \dsm{} evolved
to $X_{c}=$ 0.2607, the central temperature is 2.2978$\times 10^{7} K$ for the non-rotating model
but is 2.2966$\times 10^{7} K$ for the rotating model with $P_{0}=$ 0.5 day. The lower the central
temperature, the lower the energy product of H-burning. Thus the rotating models have slightly
lower luminosities than non-rotating ones at the same evolutionary state. As a consequence, the effective temperatures
of rotating models are lower than those of non-rotating ones.

\subsection{For the stars with mass larger than 2.1 \dsm{}}
%-------------------------------------------------------------
   \begin{figure}
   \centering
   \includegraphics[width=4cm, angle=-90]{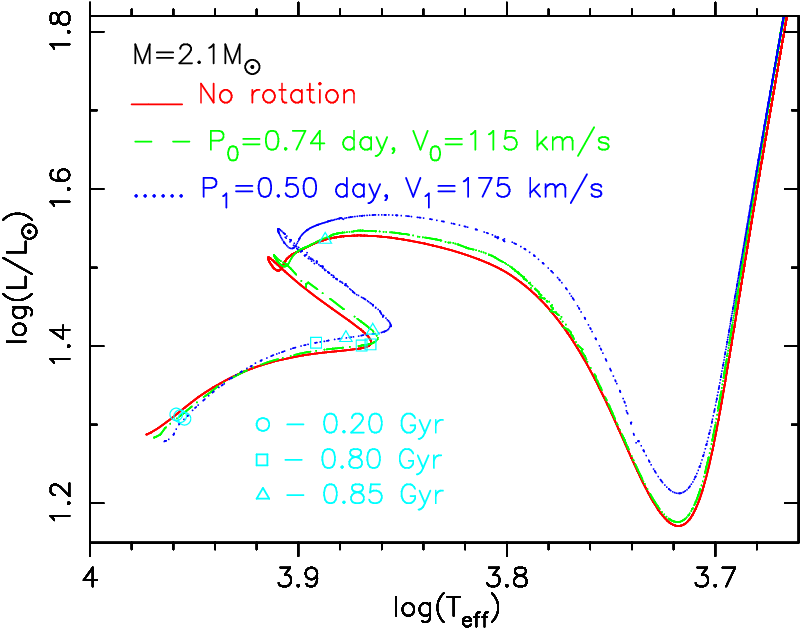}
   \includegraphics[width=4cm, angle=-90]{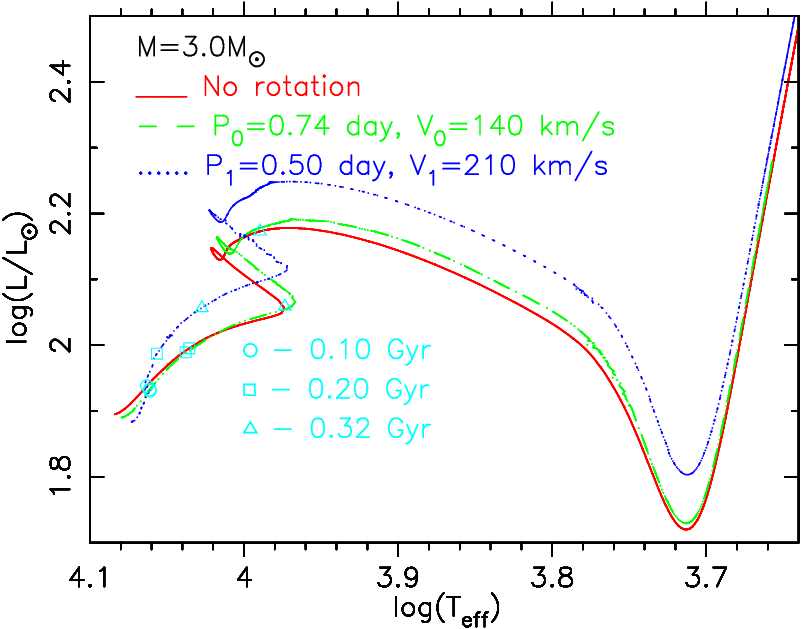}
   \caption{Same as Fig.\ref{fm1p0} but for the models with M = 2.1 and 3.0 \dsm{}, respectively.}
   \label{fig2p1}
   \end{figure}

In Fig. \ref{fig2p1}, we show the evolutionary tracks of the 2.1 and 3.0 \dsm{} models with and without rotation.
For the stars with $M \geq$ 2.1 \dsm{}, in the early stage of the MS (the $X_{c} >$ 0.5 for $M=$ 2.1 \dsm{} and
$X_{c} >$ 0.6 for $M=$ 3.0 \dsm{}) the hydrostatic effects of rotation dominate the corrections to stellar models;
thus the rotating models exhibit lower luminosities and effective temperatures than the non-rotating ones
at the same age.

As the evolution proceeds, the effects of rotation leads to an increase in the convective core for these stars,
and the more massive the star or the higher the rotation rate, the larger the change of the convective core
(see Fig. \ref{frc}). Both the increase of the convective core and rotational mixing can enhance the hydrogen
abundance and decrease the helium abundance in the core, which leads to an increase in the mean density (i.e. a decrease
in radius) and decrease in the central temperature of the rotating models compared to non-rotating ones at the same
age, and prolongs the lifetime of core H burning. In addition, the hydrostatic effects of rotation also can lead to
a decrease in the central temperature. Thus rotating models evolve slower than non-rotating ones and exhibit lower
luminosities than non-rotating ones at the same age. The effects of rotation result in an increase in the 
convective core for stars with $M \gtrsim$ 2.1 \dsm{} but to a decrease in the convective core for stars with 
$M \lesssim$ 2.0 \dsm{}. Chemical elements are mixed completely in the convective cores. The increase of the 
convective core causes that the location of the chemical element gradient moves outwards, which leads to the 
facts that the product of H-burning can be much more efficiently transported outwards by rotational mixing. 
Thus the rotationally induced element mixing in stars with
$M \gtrsim$ 2.1 \dsm{} is more efficient than that in stars with mass between about 1.7 and 2.0 \dsm{}.
For example, when the stars with $M=$ 2.0 and 2.1 \dsm{} evolved to $X_{c}=$ 0.261, the surface hydrogen
abundance is 0.707 for both non-rotating models, however it is 0.705 for the rotating
model with $M=$ 2.0 \dsm{} and $P_{0}=$ 0.5 day but is 0.700 for the rotating model with $M=$ 2.1 \dsm{} and
$P_{0}=$ 0.5 day. The efficient mixing leads to the fact that the radius of the rotating models
is smaller than that of non-rotating ones at the same age and the change of the radius is larger
that of luminosity. Thus the rotating models exhibit the higher effective temperatures than non-rotating
ones at the same age during the middle stage of MS.
However, when the models evolved into the same evolutionary state of the late stage of MS,
because the rotating models consumed more hydrogen fuel than non-rotating ones, which can enhance the temperatures
of core and the He-core mass left behind, the rotating models have a larger energy product of H burning than
non-rotating ones. The higher the energy product, the more the expansion of stars. Thus the rotating models exhibit
larger radii and lower effective temperatures than non-rotating ones at the same evolutionary state.
The effects of rotation on the evolutions of these stars
with $M\gtrsim$ 2.1 \dsm{} are similar to the effects of convective core overshooting.

\section{Discussion and conclusions}

Besides the influences of rotation on the internal evolution of stars, effects of rotation 
on the observable parameters of stars depend on the product $v\sin i$, where $v$ is the equatorial 
rotational velocity and $i$ refers to the angle between the rotational axis of a star and the 
direction towards the observor. In this work, we only focus on the effects on the internal stellar evolution.

The effects of rotation on stellar structure and evolution mainly derive from the hydrostatic
effects, the mixing of elements caused by rotationally induced instabilities, and the Von Zeipel
effect which affects the instability of convection by changing the radiative temperature gradient.
The hydrostatic effects mainly lead to an increase in radius and a decrease in the effective temperature.
The mixing of elements, however, chiefly results in an increase in the mean density, i.e. a decrease
in radius and an increase in the effective temperature. Moreover, rotation leads to
a decrease in the convective core for stars with $M <$ 2.05 \dsm{}, which can counteract the effects of
the rotational mixing and accelerate the evolution of stars. However, rotation results in an increase
in the convective core for stars with $M >$ 2.05 \dsm{}, which dominates the effect of rotation
on the evolutions of these stars. For the models with masses between about 1.7 and 2.0 \dsm{},
because the effects of the rotational mixing is counteracted by the effects of the decrease in
the convective core and they do not experience magnetic braking during the MS stage,
the hydrostatic effects dominate the effects
on the effective temperatures and luminosities of these models; thus, the effective temperatures and
luminosities of rotating models are lower than those of non-rotating ones during the MS stage.

The evolutions of our rotating models with $M\gtrsim$ 2.1 \dsm{} are consistent with the calculation results
of Eggenberger et al. (\cite{egge10}) and Girardi et al. (\cite{gira11}). However the rotating models with
mass between $\sim$1.7 and 2.0 \dsm{} manifest lower effective temperatures than non-rotating ones,
which is not consistent with the results of Girardi et al. (\cite{gira11}) but is similar to
the calculation result of Bastian \& de Mink (\cite{bast09}). Moreover,
in our models, the MS bandwidth of rotating models with $M >$ 2.0 \dsm{} is wider than that of non-rotating
ones. However, the MS bandwidth of rotating models with $M <$ 2.0 \dsm{} is narrower than that
of non-rotating ones, which is consistent with the distributions of the large sample of rotating stars
collected by Royer et al. (\cite{roye07}) in the HR diagram (see the Fig. 4 of Zorec \& Royer \cite{zore12}).
In the next work, we will give a more detailed comparison.

For MS stars with a given rotation rate, the change in the effective temperature caused by rotation
increases with increasing mass. For example, when age $=$ 1.1 Gyr and $P_{0}=$ 0.5 day,
the difference of the effective temperature between non-rotating and rotating model is about
300 K for the stars with mass between 1.7 and 1.8 \dsm{}, is around 200 K for the star with
$M=$ 1.6 \dsm{}, and is about 40 K for the star with $M=$ 1.5 \dsm{}, but is only several Kelvin
for the stars with $M\lesssim$ 1.4 \dsm{}. The change in the effective temperature caused by
rotation decreases with decreasing mass. Thus, for some intermediate-age star clusters,
rotation might lead to a color spread near the MSTO in their CMD.
In the next work, we will give and discuss the isochrone of rotating models.

In this work, we calculated a grid of evolutionary tracks of rotating models with masses between
1.0 and 3.0 \dsm{} and a resolution $\delta M \leq$ 0.02 \dsm{}.
We find that the effects of rotation on stellar structure and evolution are dependent not only
on the rotation rate but also on the mass of stars. For stars with $M >$ 2.05 \dsm{}, rotation leads to
an increase in the convective core and prolongs the lifetime of core H burning; the evolution of rotating models
of these stars is slower than that of non-rotating ones; in the early stage of MS, the changes in luminosities
and effective temperatures are mainly due to the hydrostatic effects of rotation, thus rotating models
exhibit lower luminosities and effective temperatures than non-rotating ones at the same age; however,
in the late stage of MS, rotating model can manifest higher effective temperatures than non-rotating ones
at the same age and larger luminosities at the same evolutionary stage because the rotational mixing dominates
the effects on the models. For stars with 1.1 \dsm{} $\lesssim M <$ 2.05 \dsm{}, rotation results in
a decrease in the convective core and shortens the lifetime of core H burning; the rotating models of these stars
evolve faster than non-rotating ones. When 1.7 \dsm{} $\lesssim M <$ 2.05 \dsm{}, the rotating models exhibit
lower effective temperatures but approximately equal luminosities compared to non-rotating ones at the same age;
the evolutionary tracks of the rotating models are located on the lower right of that of
non-rotating ones in the HR diagram, which are mainly due to the hydrostatic effects of rotation and `rotation
acceleration'. However,
for stars with 1.0 \dsm{} $\lesssim M\lesssim$ 1.4 \dsm{}, due to the fact that these stars experienced
the magnetic braking from the beginning of evolution, the rotational mixing and `rotation acceleration'
dominate the effects of rotation on the evolution of these stars; the evolutionary tracks of rotating models
of these stars are mainly located on the left of that of non-rotating ones in the HR diagram;
the rotating models can exhibit lower or higher effective temperatures than non-rotating ones at the same age,
which depends on the mass and age of stars. Our calculations show that the mass of 2.05 \dsm{} is a critical value
for the effect of rotation on the stellar structure and evolution. This value is very close to the critical 
mass (2.01 \dsm{}) for oscillations of horizontal branch stars (Yang et al. \cite{yang12}). 
Rotation could lead to a color spread for some intermediate-age star clusters near the MSTO in their CMD.

\begin{acknowledgements}
This work was supported by China Postdoctoral Science Foundation through grant 20100480222, 
the NSFC through grants 11273012, 11273007, 10773003, 10933002, and 11003003, the Ministry of 
Science and Technology of the People's republic of China through grant 2007CB815406, and
the Project of Science and Technology from the Ministry of Education (211102).
\end{acknowledgements}

\label{lastpage}

\end{document}